\documentclass[aps,prl,twocolumn,showpacs,superscriptaddress]{revtex4-1}
\usepackage{bm}
\usepackage{mathrsfs}
\usepackage{amsmath}
\usepackage{amssymb}
\usepackage{graphicx}
\usepackage{amsfonts}
\usepackage{amsthm}
\usepackage{color}
\usepackage{dcolumn}
\usepackage{txfonts}

\begin{document}

\title{Magnetic ordering of nitrogen-vacancy centers in diamond via resonator-mediated coupling}
\author{Bo-Bo Wei}
\affiliation{Department of Physics, The Chinese University of Hong Kong, Hong Kong, China}
\affiliation{Center for Quantum Coherence, The Chinese University of Hong Kong, Hong Kong, China}
\author{Christian Burk}
\affiliation{3rd Institute of Physics and Research Center SCOPE, University Stuttgart, Stuttgart 70569, Germany}
\author{J\"{o}rg Wrachtrup}
\affiliation{3rd Institute of Physics and Research Center SCOPE, University Stuttgart, Stuttgart 70569, Germany}
\author{Ren-Bao Liu}
\email{Corresponding author. rbliu@phy.cuhk.edu.hk}
\affiliation{Department of Physics, The Chinese University of Hong Kong, Hong Kong, China}
\affiliation{Center for Quantum Coherence, The Chinese University of Hong Kong, Hong Kong, China}
\affiliation{Institute of Theoretical Physics, The Chinese University of Hong Kong, Hong Kong, China}
\affiliation{Shenzhen Research Insitute, The Chinese University of Hong Kong, Shenzhen, Guangdong 518057, China}

\begin{abstract}
Nitrogen-vacancy centers in diamond, being a promising candidate for quantum information processing, may also be an ideal platform for simulating many-body physics.  However, it is difficult to realize interactions between nitrogen-vacancy centers strong enough to form a macroscopically ordered phase under realistic temperatures. Here we propose a scheme to realize long-range ferromagnetic Ising interactions between distant nitrogen-vacancy centers by using a mechanical resonator as a medium. Since the critical temperature in the long-range Ising model is proportional to the number of spins, a ferromagnetic order can be formed at a temperature of tens of millikelvin for a sample with $\sim10^4$ nitrogen-vacancy centers. This method may provide a new platform for studying many-body physics using qubit systems.
\end{abstract}


\maketitle
\emph{Introduction.}
The negatively charged nitrogen-vacancy (NV) centers in high-purity
diamond have been considered as a promising candidate
for solid state quantum information processing due to their long
coherence time \cite{Kennedy2003,Wrachtrup2006,Wrachtrup2009}
and high feasibility in initialization, control, and readout of their spin states \cite{Wrachtrup1997}.
Simulation of many-body physics using NV centers, in analogue to cold atom physics \cite{Lewenstein2007,Bloch2008}, has been proposed \cite{Lukin2010,Lukin2012,Cai2012,Chen2013}. To realize phase transitions in NV center qubit systems under realistic temperatures, however, sufficiently strong interactions between NV centers located sufficiently close ($<$30 nm) is till highly challenging\cite{Lukin2010,Lukin2012,Wrachtup2013a}. A new opportunity is to use resonators as mediators \cite{Lukin2010,Lukin2009,Du2009}, which has potential of coupling NV centers at distance.

In this letter, we propose to realize long-range coupling between many separated NV centers via a mechanical resonator. A remarkable feature of the long-range interacting system is: the critical temperature for ferromagnetic phase transitions is proportional to the number of spins, so a ferromagnetic order could be formed at a temperature of tens of millikelvin (mK) for a sample with $\sim 10^{4}$ NV centers, even though the mediated coupling between two NV centers is less than 200 kHz($\sim 10\ \mu$K).

\begin{figure}
\begin{center}
\includegraphics[width=\columnwidth]{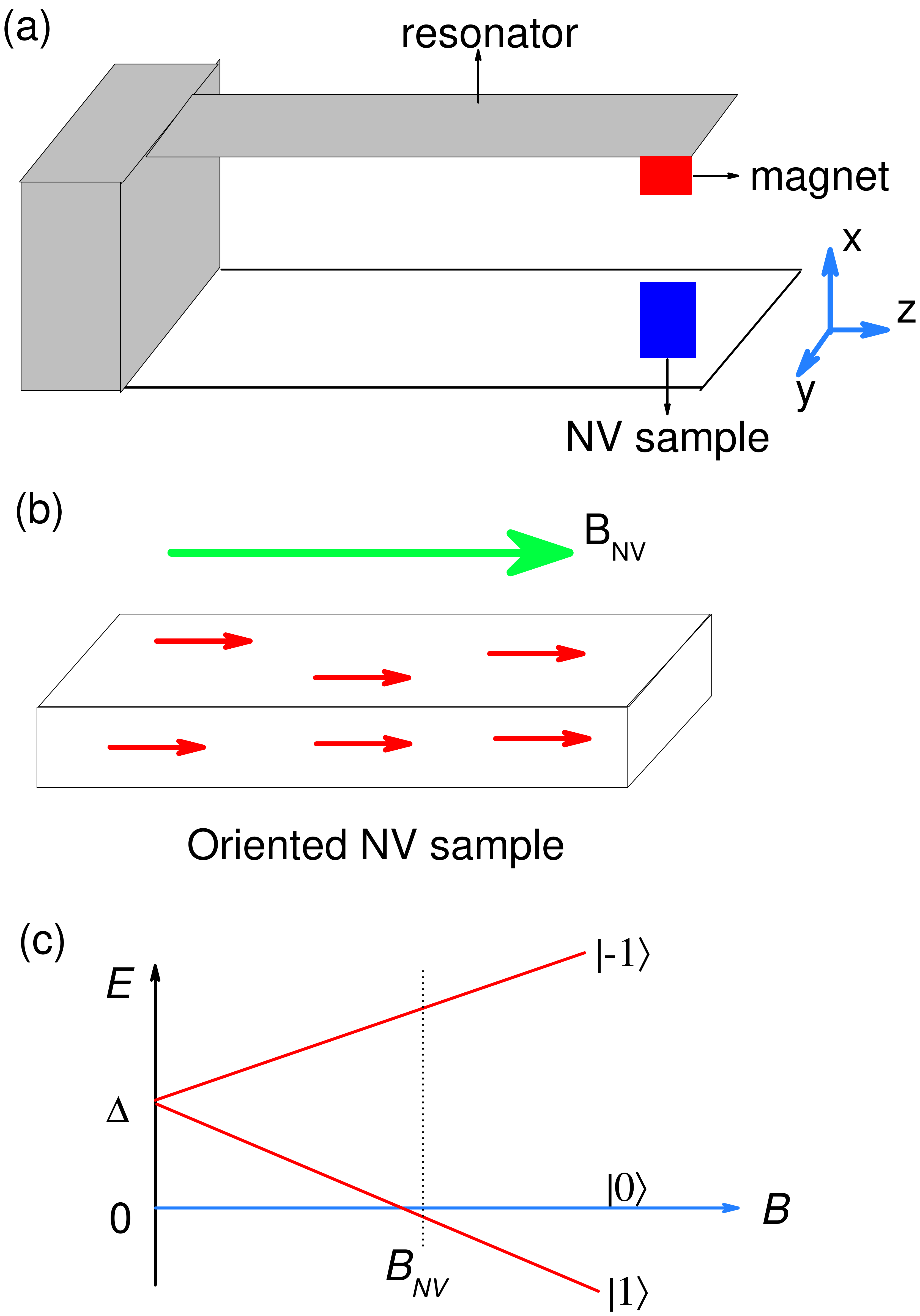}
\end{center}
\caption{(color online).  (a). Schematic set up of a hybrid NV and mechanical resonator system. A magnet attached to the end of the mechanical resonator is positioned at a
distance 25 nm above the NV centers, coupled to the electronic spin of the defect centers. (b). The NV defects in the samples. The defects have one preferred orientations in the sample, $[111]$. A static magnetic field $B_{NV}$ is applied along the $[111]$ direction. (c). Schematic energy levels of NV centers as a function of the magnetic field. The $B_{NV}$ is taken so that the states $|1\rangle$ and $|0\rangle$ of the NV centers are nearly degenerate.}
\label{fig:epsart1}
\end{figure}

\emph{Microscopic model}.
The essential idea can be understood
by considering a prototype system shown in Fig.~\ref{fig:epsart1}(a). A mechanical resonator with frequency $\omega_r$ is attached with a magnet. The NV sample is placed right under the magnet with a distance $d$. The oscillation of the mechanical resonator generates a time dependent magnetic field that causes Zeeman shift of the spins in the sample. In contrast to the schemes of direct dipolar coupling between NV centers, our scheme does not require short distance between NV centers. We use a diamond sample where the orientation axes of the NV color centers can lie along one of the four possible crystallographic axes in diamonds, $[111]$ (Fig.~\ref{fig:epsart1}(b)) which can be grown by chemical vapor deposition method \cite{Fu2012,Pham2012}. The NV centers have a spin triplet ground state ($S=1$) with a large zero field splitting $\Delta=2.87$ GHz. We apply a static magnetic field $B_{NV}$ along the $[111]$ direction of the sample so that two Zeeman states of the electron spin qubits with magnetic quantum number $|1\rangle$ and $|0\rangle$ are nearly degenerate, $\delta\equiv(\Delta-\gamma_eB_{NV})\sim0$ (Fig.~\ref{fig:epsart1}(c)).

We make use of the two Zeeman states of the electron spin $|1\rangle$ and $|0\rangle$ and define a pseudo-spin by $\sigma_z=|1\rangle\langle1|-|0\rangle\langle0|$ and the corresponding spin flip operators $\sigma^{\dagger}=|1\rangle\langle0|$ and $\sigma^{-}=|0\rangle\langle1|$. In each NV center there is a $^{14}\text{N}$ nuclear spin which interacts with the on-site NV center spin. The Hamiltonian of the NV center near the degenerate point ($\delta\approx 0$) is $H_{\text{NV}}=\sum_j[\Delta_N(I_j^z)^2-\gamma_NI_j^z B_{NV}+A I_j\cdot \sigma_j+\delta\sigma_j^z]$, where $\Delta_N=5.1$ MHz is the $^{14}\text{N}$ nuclear spin quadrupole splitting \cite{He1993}, $\gamma_N$ is the gyromagnetic ratio of the nitrogen nuclear spin and $A\approx 2$ MHz is the hyperfine coupling between the electronic spin and the $^{14}\text{N}$ nuclear spin \cite{He1993}.

The motion of the mechanical resonator is described by the Hamiltonian $H_r=\omega_rb^{\dagger}b$, with
$\omega_r$ as the fundamental vibration mode of the resonator and $b$
as the corresponding annihilation operator. For example a silicon nitride
string resonator has dimensions $(325\times 0.35\times0.1 )\mu$m with $\omega_r=2\pi\times 1.0$ MHz and $Q =1.3\times10^6$ at room temperature \cite{Verbridge2008}.

The magnetic field felt by the electronic spin of the NV centers
can be approximated by a magnetic dipole \cite{Hunger2011}, $\vec{B}(d_j-x(t))=\vec{B}(d_j)-G_m\vec{x}(t)+o(x^2)$, where $d_j$ is the distance between the equilibrium position of the resonator and $j-$th NV center, $G_m$ is the magnetic field gradient at the position of the NV centers and $x$ is the amplitude of mechanical resonator oscillation ($x\sim 10^{-12}$ m at temperature of mK). This magnetic field will induce a Zeeman shift to the NV spins with Hamiltonian
$H_z=\sum_j\eta_j\sigma_j^x(b+b^{\dagger})$ and $\eta_j=\gamma_e|G_{m}|a_0$, where $\gamma_e$ is the gyromagnetic ratio for electron and $a_0=\sqrt{\hbar/2m\omega_r}$ is the amplitude of zero point fluctuations
for a resonator of mass $m$.

Due to the distance distribution of the NV centers to the magnet, the coupling of the NV centers to the mechanical resonator have a distribution. 
A magnetic tip with size of 100 nm produces a magnetic gradient $G_m\sim 7.8\times 10^6$ T/m at a distance 25 nm away from the tip \cite{mamin2007}.
 In such case $\eta/2\pi$ could reach 200 kHz.  For a sample with typical distance of NV centers $\sim 20$ nm, the neighbor NV centers have a direct dipolar interaction with strength $\sim 5$ kHz ($\sim 0.25 \ \mu$K), much less than the critical temperature to be discussed later. For the sake of simplicity, in the following we assume the coupling between the mechanical resonator and NV centers to be uniform and neglect the direct dipolar coupling between NV centers. Without these assumptions, however, the results in this paper would only be quantitatively affected. The coupling constant $\eta/2\pi\sim 200$ kHz considerably exceeds
both the electronic spin coherence time (milliseconds) of the NV centers and the
intrinsic damping rate of the mechanical resonator, $\gamma=\omega_r/Q$ of high-$Q$ mechanical resonator \cite{Verbridge2008}. We neglect the interaction between the mechanical resonator and the $^{14}\text{N}$ nuclear spins since they are three orders of magnitude smaller than that between the NV center and the mechanical resonator.

Summarizing the three terms
described above, the total microscopic Hamiltonian including the spin qubits in the NV centers, nitrogen nuclear spins,
the mechanical resonator and the interaction among them is,
\begin{eqnarray}
H&=&\omega_rb^{\dagger}b+\sum_{j=1}^N\Big[\Delta_N(I_j^z)^2-\gamma_NB_{NV}I_j^z +A I_j\cdot \sigma_j\Big]\nonumber\\
& & +\sum_{j=1}^N\Big[\eta(b+b^{\dagger})\sigma_{j}^x+\delta\sigma_j^z\Big].
\end{eqnarray}

\begin{figure}
\begin{center}
\includegraphics[width=\columnwidth]{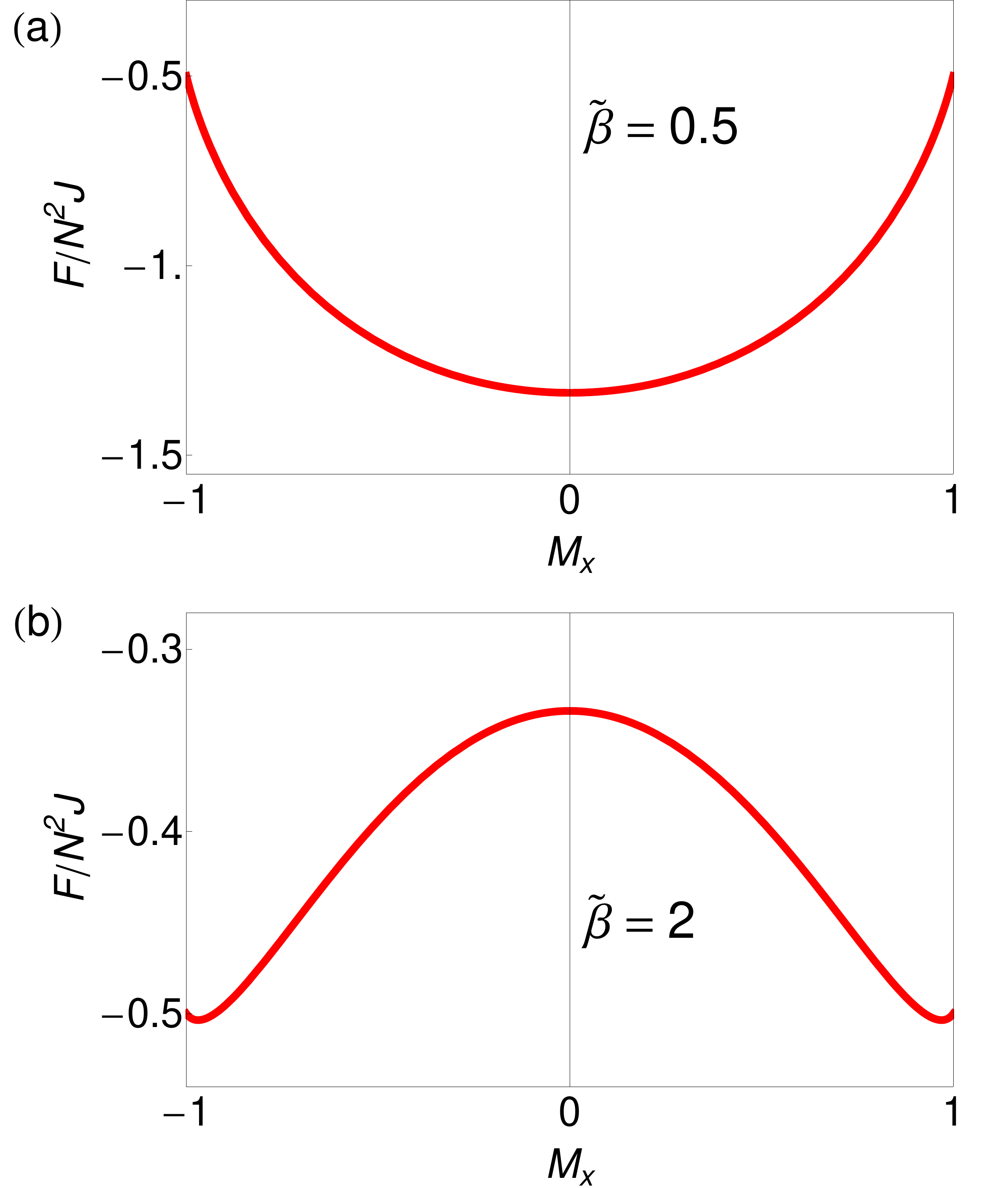}
\end{center}
\caption{(color online). Free energy landscapes of the long-range Ising model. (a). The free energy as a function of the spin polarization along the $x$ direction at high temperature $\tilde{\beta}\equiv NJ/k_BT=0.5$. (b). The same as (a) but at a lower temperature $\tilde{\beta}\equiv NJ/k_BT=2.0$. }
\label{fig:epsart2}
\end{figure}

\emph{Effective long-range Ising model.}
To obtain the effective interactions between NV centers, we calculate the partition function of the hybrid system $Z=\text{Tr}[e^{-\beta H}]$.
We shall neglect the Zeeman energy of the $^{14}\text{N}$ nuclear spin since its strength ($\sim0.3$ MHz) is much smaller than the zero field splitting $\sim 5$ MHz and the hyperfine interaction $\sim 3$ MHz. The hyperfine interaction between the electronic spin and the $^{14}\text{N}$ nuclear spin term $AI\cdot\sigma$ can be separated into the diagonal term $AI^x\sigma^x$ and the flip-flop term $A(I^y\sigma^y+I^z\sigma^z)$. Let us first neglect the flip-flop term and the qubit splitting ($\delta=0$). By partial trace of the phonon bath and the nuclear spins, the partition function can be factorised as
\begin{eqnarray}
Z=\text{Tr}[e^{-\beta H}]=Z_{\text{phonon}}\times Z_{^{14}\text{N}}\times \text{Tr}[\exp(-\beta H_{\text{eff}})],
\end{eqnarray}
where $Z_{\text{phonon}}$ is the partition function of an independent phonon
bath, $Z_{^{14}\text{N}}$ is the partition function of the nuclear spins and the effective Hamiltonian for the NV electron spins is
\begin{eqnarray}
H_{\text{eff}}=-\sum_{i<j}\frac{2\eta^2}{\omega_r}\sigma_i^x\sigma_j^x-\beta h(\beta)\sum_j\sigma_j^x,
\end{eqnarray}
with $h(\beta)=\beta^{-1}\ln[(1+2\cosh(\beta A_x))/3]$ being an effective magnetic field resulting from the on-site interaction between the NV center and the $^{14}$N nuclear spin. The physics for the effective long-range
interactions between the NV centers is that an NV center spin flips by virtually emitting a phonon, and then the phonon is virtually absorbed by another NV center spin far away with a spin flip. Therefore two NV spins at distance can flip-flop via virtual exchange of phonons, which leads to long-range $\sigma_i^x\sigma_j^x$ coupling. In the temperature range we are interested in($\sim$ 10 mK), the effective field, $h(\beta)\sim0.3 \mu$K, is negligible. Therefore the distant NV centers interact through an effective long-range ferromagnetic Ising interaction with the Hamiltonian
\begin{eqnarray}\label{model}
H_{\text{electron}}=-\sum_{i<j}J\sigma_i^x\sigma_j^x,
\end{eqnarray}
where the ferromagnetic coupling strength $J=2\eta^2/\omega_r$. If the flip-flop term of the hyperfine interaction is taken into account, it would contribute an effective transverse field to the electronic spins, which, however, is $\sim0.3 \mu$K and hence negligible.

\begin{figure}
\begin{center}
\includegraphics[width=\columnwidth]{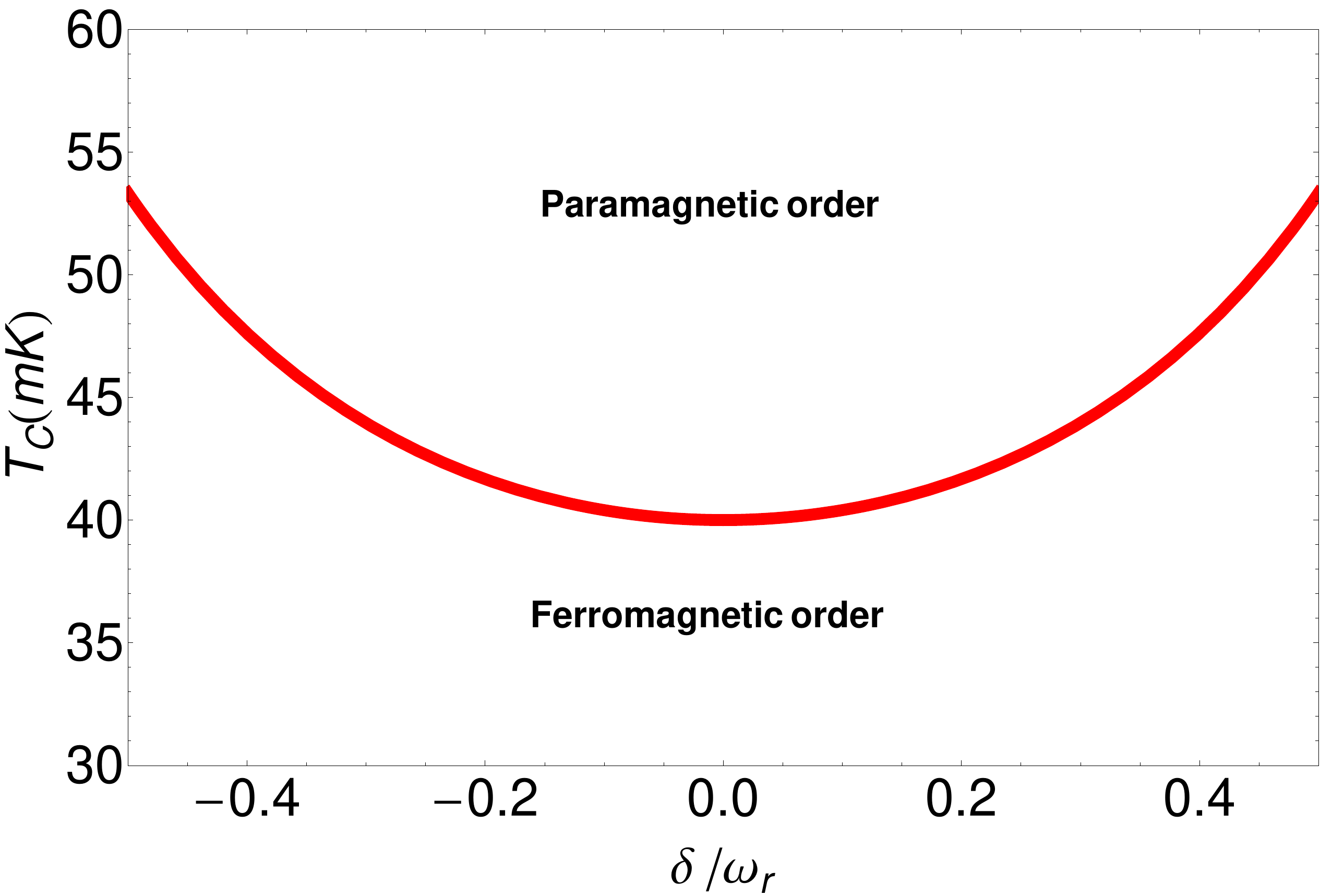}
\end{center}
\caption{(color online). The phase transition temperature of the NV-resonator hybrid system as a function of the scaled detuning $\delta/\omega_r$. We take the NV-mechanical resonator interaction strength $\eta=200$ kHz, eigen frequency of the mechanical resonator $\omega_r=1$ MHz and the number of spin is $N=10000$. }
\label{fig:epsart3}
\end{figure}

\begin{figure}
\begin{center}
\includegraphics[width=\columnwidth]{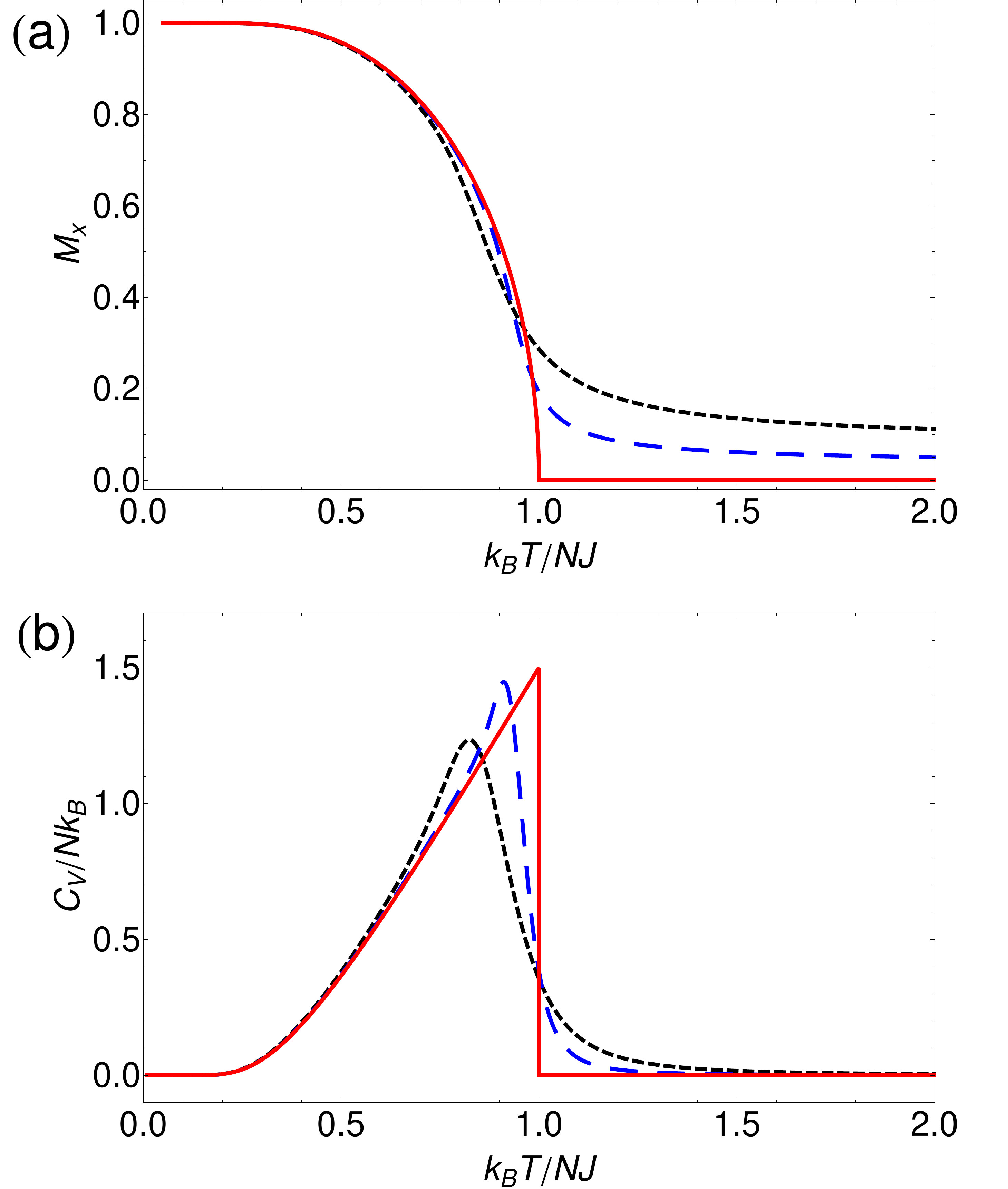}
\end{center}
\caption{(color online). Comparison of the finite size numerical results and the mean field theory. (a). The order parameter, $M_x$, as a function of
the reduced dimensionless temperature, $k_BT/NJ$ for different sizes of the spin systems. The black-short-dashed line is for $N=100$, the blue-long-dashed line is for $N=500$ and
the red-solid line is the mean field prediction. (b). The specific heat plot in the same style as in (a). }
\label{fig:epsart4}
\end{figure}

\emph{Ferromagnetic ordering.}
The long-range Ising model can be solved by mapping the ensemble onto to a single large spin.
With the notation $S_x=\sum_{j=1}^N\sigma_j^x/2$, the Hamiltonian of the long-range Ising model can be written as
\begin{eqnarray}
H_{\text{electron}}=-J\sum_{i<j}\sigma_i^x\sigma_j^x=-J\Big(2S_x^2-\frac{N}{2}\Big).
\end{eqnarray}
The spin quantum number $S$ takes values $0,1,2,\cdots, N/2$ for even $N$ and $1/2,3/2,\cdots,N/2$ for odd $N$. Moreover,
the spin degeneracy in each large spin subspace is $D[S]=C_{N}^{N/2-S}-C_N^{N/2-S-1}$ where $C_N^j$ is the binomial coefficient. Then the electron spin partition function
\begin{eqnarray}
Z=\text{Tr}[e^{-\beta H_{\text{electron}}}]=e^{-N\beta J/2}\sum_{n=0}^NC_N^ne^{2\beta JN^2(n/N-1/2)^2}.
\end{eqnarray}
In the large $N$ limit \cite{Wilms2012}
\begin{eqnarray}
Z\approx e^{-\tilde{\beta}/2}N\int_{-1/2}^{1/2}dX e^{-N\varphi(\tilde{\beta},X)},
\end{eqnarray}
where $\varphi=(1/2+X)\ln(1/2+X)+(1/2-X)\ln(1/2-X)-2\tilde{\beta}X^2$ and $\tilde{\beta}=\beta NJ$. The integration in the partition function can be evaluated by the saddle point approximation method. For $\tilde{\beta}<1$, the saddle point appears at $X=0$, and for $\tilde{\beta}>1$, there are two symmetric saddle point located respectively within $(-1/2,0)$ and $(0,1/2)$. Therefore $\tilde{\beta}=1$  is the critical point for the ferromagnetic phase transition in the long-range Ising model. So the critical temperature for the long-range ferromagnetic Ising model, $k_BT_c=NJ$ , is proportional to the number of spins and the
coupling strength \cite{Huang1987}.

We present the scaled free energy of the long-range Ising model as a function of the spin magnetization $M_x\equiv\langle S_x\rangle/N$ for different temperatures in Fig.~\ref{fig:epsart2}.  One can see that at high temperature (Fig.~\ref{fig:epsart2}(a)) the free energy minimum is situated at $M_x=0$ and a paramagnetic state is stable, while at low temperature (Fig.~\ref{fig:epsart2}(b)) the free energy minimum is doubly degenerated with $M_x=1$ or $M_x=-1$, and a ferromagnetic state is preferred.

For the hybrid system, the phase transition temperature can be controlled through the detuning from the degeneracy point between the NV spin states $|1\rangle$ and $|0\rangle$. If the NV qubits have a small splitting $\delta=\Delta-\gamma B_{NV}$, the resonator mediated interaction between the NV centers becomes
\begin{eqnarray}
J=-\Big(\frac{\eta^2}{\omega_r-\delta}+\frac{\eta^2}{\omega_r+\delta}\Big),
\end{eqnarray}
The critical temperature becomes
\begin{eqnarray}
k_BT_c=\frac{N\eta^2}{\omega_r-\delta}+\frac{N\eta^2}{\omega_r+\delta}.
\end{eqnarray}
Here the magnetic field is tuned so that the NV spin states $|1\rangle$ and $|0\rangle$ are near degenerate. Hence the splitting $\delta$ is much smaller than the critical temperature. In Fig.~\ref{fig:epsart3} we plot the phase diagrams of the NV centers as a function of splitting for $N=10000$. In a diamond sample with NV concentration $\sim 50$ ppb, coupling between a resonator and such many NV centers is possible. A ferromagnetic order can be formed at the temperature of 50 mK.

In Fig.~\ref{fig:epsart4}(a), we present the numerical and the mean field results of the order parameter as a function of reduced temperature, $k_BT/NJ$. One can see that the order parameter vanishes at and above the critical temperature $k_BT/NJ=1$ and also the finite size numerical results approach to that of the mean field predictions reasonably well as the number of spins increases. The specific heat shown in Fig.~\ref{fig:epsart4}(b) has a finite jump which signals a second order phase transitions occurs.

\emph{Conclusion.} In summary, we propose a scheme to realize ferromagnetic ordering of distant nitrogen-vacancy centers by using a mechanical resonator to mediate long-range Ising-type interaction. The critical temperature for the ferromagnetic phase transition in the long-range Ising model is proportional to the number of spins, so the ferromagnetic order could be formed at the temperature of tens of millikelvin for a sample with ten thousand nitrogen-vacancy centers. In addition, it may also be possible to use a superconducting resonator \cite{Zhu2011,Kubo2011} as a medium to realize the long-range ferromagnetic coupling between the NV centers. Since the interactions between the NV centers mediated by a superconducting resonator is usually small compared to that by a mechanical resonator, high density NV samples are required to observe the magnetic ordering in the NV centers.

\begin{acknowledgements}
We are grateful to P. Bertet for useful discussions. This work was supported by Hong Kong Research Grants Council/General
Research Fund CUHK402410, The Chinese University of Hong Kong
Focused Investments Scheme, Hong Kong Research Grants
Council/Collaborative Research Fund HKU8/CRF/11G, and National Basic Research Program of China Grant 2014CB921402.
\end{acknowledgements}

\end{document}